\documentclass[%
aps,
prl,
twocolumn,
superscriptaddress,
amsmath,
amssymb,
citeautoscript
]{revtex4-1}
\usepackage{graphicx}
\usepackage{dcolumn}
\usepackage{bm}
\usepackage{hyperref}
\usepackage{float}
\usepackage{xcolor}
\usepackage{soul}
\usepackage[normalem]{ulem}

\newcommand{\RNum}[1]{\uppercase\expandafter{\romannumeral #1\relax}}

\newcommand{\siexampleturbulence}{S1}

\newcommand{\siexampleanchoring}{S4}

\begin{document}

\preprint{APS/123-QED}

\title{
Why Extensile and Contractile Tissues Could be Hard to Tell Apart
}

\author{Jan Rozman}
\email{jan.rozman@ijs.si}
\affiliation{Rudolf Peierls Centre for Theoretical Physics, University of Oxford, Oxford OX1 3PU, United Kingdom}
\affiliation{Faculty of Mathematics and Physics, University of Ljubljana, Jadranska 19, SI-1000 Ljubljana, Slovenia}
\affiliation{ Jo\v zef Stefan Institute, Jamova 39, SI-1000 Ljubljana, Slovenia}

\author{Sumesh P. Thampi}%
\affiliation{Rudolf Peierls Centre for Theoretical Physics, University of Oxford, Oxford OX1 3PU, United Kingdom}
\affiliation{Department of Chemical Engineering, Indian Institute of Technology, Madras, Chennai, India 600036}

\author{Julia M. Yeomans}%
\affiliation{Rudolf Peierls Centre for Theoretical Physics, University of Oxford, Oxford OX1 3PU, United Kingdom}
\date{\today}

\begin{abstract}
Active nematic models explain the topological defects and flow patterns observed in epithelial tissues, but the nature of active stress---whether it is extensile or contractile, a key parameter of the theory---is not well established experimentally. Individual cells are contractile, yet tissue-level behavior often resembles extensile nematics. To address this discrepancy, we use a continuum theory with two-tensor order parameters that distinguishes cell shape from active stress. We show that correlating cell shape and flow, whether in coherent flows in channels, near topological defects, or at rigid boundaries, cannot unambiguously determine the type of active stress. Our results demonstrate that simultaneous measurements of stress and cell shape  are essential to fully interpret experiments investigating the  nature of the physical forces acting within epithelial cell layers.
\end{abstract}

\maketitle


{\textit{Introduction}}---The theory of active nematics provides a generic framework for understanding how epithelial cell layers and tissues move and organize collectively. This approach has been useful in explaining both disordered ~\cite{saw2017topological,blanch2018turbulent,balasubramaniam2021investigating} and coherent~\cite{duclos2018spontaneous, serrano2025control} cellular flows. The theory holds when there is no net propulsive motion of the tissue. Because the forces are balanced, they are described to leading order as force dipoles,  pairs of equal and opposite forces, which can be coarse-grained and represented as an active stress in continuum theories.

An active nematic is classified as \textit{contractile} if the forces of the dipole pull inwards along the dipolar axis. Similarly, an active nematic is termed \textit{extensile} if the forces act outwards~\cite{ramaswamy2010mechanics}. The nature of active stress in  active nematic systems, such as reconstituted cytoskeletal extracts powered by molecular motors~\cite{sanchez2012spontaneous,shelley2016dynamics, doostmohammadi2018active} and microswimmer suspensions~\cite{qi2022emergence,ishikawa2024fluid}, has been characterized to a significant extent through theory and experiments. However, such progress remains more limited in the context of cell assemblies and tissues which are deformable, and which can lack nematic symmetry in the passive limit.
 
While individual cells are expected to be contractile---owing to actomyosin-mediated contractility---experimental observations of monolayer dynamics often align more closely with active nematic models with extensile stress~\cite{saw2017topological,duclos2018spontaneous}. Several authors \cite{zhang2023active, balasubramaniam2021investigating, vafa2021fluctuations, killeen2022polar} have proposed explanations for this dilemma, but the nature of activity in cell layers, and how to model this within a continuum active nematic approach, is still not fully understood.  Therefore, in this Letter we  study and characterize cellular active nematic flows in confinement. Our results illustrate why identifying the extensile–contractile distinction in experiments is, unexpectedly, nontrivial for epithelial cell layers.

Active nematic theories of epithelia generally assume that the nematic active stress aligns along the long axis of the cells so that the dynamics can be described by a single tensor order parameter. Recent experiments and theories have shown that this may not be the case, developing a description distinguishing cell shape from the nematic activity field~\cite{nejad2024stress,nejad2025cellular}. Applying such a two tensor model to study confined active nematics, we find a plausible explanation for the contractile-extensile discrepancy found in the literature. We focus on three cases: i) coherent channel flows, ii) the dancing state of topological defects, and iii) cell orientation at rigid walls. In all three, we find that, irrespective of the type of activity, the analysis of the orientation of cell shapes within the framework of a single-tensor active nematic theory would suggest that the system is extensile. 

\begin{figure}[b]
    \includegraphics[width=8.2cm]{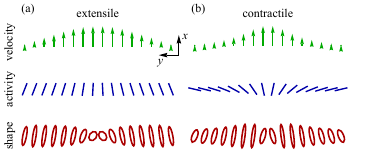}
    \caption{
    Top to bottom: velocity, active stress director, and cell shape profiles along a cross-section of the channel during unidirectional flow for (a) an extensile and (b) a contractile cellular continuum. Note the coordinate system is always oriented so that the $x$ axis corresponds to the direction of flow.
    }
    \label{fig:profiles}
\end{figure}

\begin{figure}[h]
    \includegraphics[width=8.2cm]{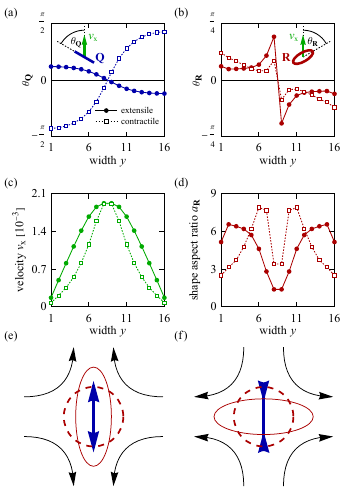}
    \caption{
    Comparison of channel flows in an extensile and a contractile system: profiles of (a) $\theta_{\mathbf{Q}}$, the angle between the active stress director and the flow direction, (b) $\theta_{\mathbf{R}}$, the angle between the long axis of the shape and the flow direction, (c) velocity, and (d) shape aspect ratio. Insets of panels (a) and (b) illustrate the definitions of the angles $\theta_{\mathbf{Q}}$ and $\theta_{\mathbf{R}}$: they are the smallest angle between flow direction (channel axis) and the director/shape long axis, defined such that the value of the angle increases as the director/shape is rotated counterclockwise. That is, the inset of (a) shows a positive angle and the inset of (b) a negative angle. (e,f) Schematic comparing cell elongation in extensile and contractile systems: (e) extensile active flows (black curved arrows) extend an initially isotropic cell (dashed red circle to ellipse) along the active stress director (double headed blue arrow), whereas (f) contractile flows extend the cell perpendicular to the active stress director.
    }
    \label{fig:comparison}
\end{figure}

\textit{The model}---We present a two-tensor continuum theory to model the dynamics of cellular layers. Unlike conventional active nematic \cite{doostmohammadi2018active} or active gel \cite{marchetti2013hydrodynamics} frameworks, here the microstructure of the epithelia is described by two order parameters: a shape tensor field $\mathbf{R}$ that encodes the elongation and orientation of the cell shape and a stress tensor field $\mathbf{Q}$ that describes the nematic activity \cite{nejad2024stress,nejad2025cellular}. The evolution of the cell shape and of the active stress, in the cellular continuum, are 
\begin{align}
\partial_t \mathbf{W} + \mathbf{u}\cdot \boldsymbol{\nabla} \mathbf{W} - \mathbf{S}_{\mathbf{W}} = \Gamma_{\mathbf{W}} \mathbf{H}_{\mathbf{W}},\label{eq:Wdynamics}
\end{align} 
for $\mathbf{W} \in \{\mathbf{Q}, \mathbf{R}\}$, where the LHS represents the upper convected derivative for the second rank tensor $\mathbf{W}$ and $\Gamma_{\mathbf{W}}$ determines the relaxation time for the molecular potential $\mathbf{H}_{\mathbf{W}}$. The flow dynamics is described by a velocity field $\mathbf{u}$ and follows the incompressible Navier-Stokes equations:
\begin{align}
\rho \left(\partial_t + \mathbf{u}\cdot\nabla\right) \mathbf{u} = \nabla\cdot\boldsymbol{\Pi}; \quad\quad \nabla\cdot\mathbf{u} &=0.\label{eq:nstogether}
\end{align}
Here $\rho$ is the material density and $\boldsymbol{\Pi}$ is the stress field accounting for (i) the viscous stresses, (ii) the elastic stresses arising from the deformations in the cell shape $\mathbf{R}$ and in the nematic order $\mathbf{Q}$, and (iii) the active stress $-\zeta\mathbf{Q}$. In the third contribution, as in conventional active nematic theory, $\zeta < 0$ and $\zeta > 0$ represent contractile and extensile activity, respectively~\cite{doostmohammadi2018active, doostmohammadi2021physics}. 

Full details of Eqs.~\eqref{eq:Wdynamics} - \eqref{eq:nstogether} are given in Appendix A. A summary of the important physics of this two tensor framework is as follows: the field $\mathbf{Q}$, which represents the local active stress generated in the cellular layer, allows us to calculate the nematic axis of the local dipolar forces. The active stress induces flows and flow gradients which feed back on $\mathbf{Q}$, but also stretch and re-orient the cells as described by the shape tensor $\mathbf{R}$. From $\mathbf{R}$ we obtain the nematic shape axis and aspect ratio of the deformed cells.

The governing equations are solved numerically using a hybrid lattice Boltzmann method \cite{marenduzzo2007steady, thampi2014vorticity} in two dimensions with parameters, typically used in the active nematic literature, listed in Appendix B together with the details of the numerics. However we expect our conclusions to be broadly independent of the details of the two-tensor theory. 


{\textit{Unidirectional flows}}---Cells often migrate collectively along confined pathways shaped by their surrounding microenvironment, e.g.~during cancer progression \cite{comba2022spatiotemporal} or embryonic development \cite{lecuit2007cell, heisenberg2013forces}, and several experiments have studied unidirectional cellular flows \cite{lacroix2024emergence, duclos2018spontaneous, chepizhko2025confined}. Therefore, we first consider a layer of cells confined in a channel, modeled using the two tensor theory, distinguishing the cell shape and active stress fields. Boundary conditions at the channel walls are no-slip for velocity $\mathbf{u}$, no  anchoring for the $\mathbf{Q}$ field, and Neumann conditions for the shape field $\mathbf{R}$. 

Spontaneous flow transitions are a characteristic of active fluids~\cite{voituriez2005spontaneous,thampi2022channel, lavi2025nonlinear}; we find that when active stress exceeds a critical value, unidirectional, channel-wide flows are produced for both extensile and contractile activities, as in the case of single-tensor active nematics. The resulting velocity, active stress director, and shape profiles for our model during unidirectional flow are shown in Fig.~\ref{fig:profiles} and Fig.~\ref{fig:comparison}(a)-(d). The active stress director profile along the cross-section of the channel forms a `splay' for both extensile and contractile activity.  Specifically, the converging (diverging) end of the splay is in the direction of flow for extensile (contractile) activity, in agreement with previous investigations~\cite{duclos2018spontaneous, edwards2009spontaneous,marenduzzo2007steady}. The  cells deform under the unidirectional flow, and the resulting elongated cells form a splay configuration similar to that of the active stress director field, but with an important distinction: the converging end of the splay points in the direction of the flow in both extensile and contractile systems [Figs. 1 and 2(a),(b)]. 

Thus, it is difficult to determine whether the system is extensile or contractile solely by analyzing the two commonly observed experimental quantities, the flow field and the cell shape, in channel-confined cellular flows. This is because these two fields appear qualitatively similar for both types of activity. Moreover if, as is commonly the case, the cell shape is taken as a proxy for the active stress field, one would conclude that the cellular layer has extensile activity even in a contractile system. This interpretation arises because, for an extensile system in the classic single tensor active nematic theory, the converging end of the splay deformation in the active stress director field orients in the direction of flow \cite{edwards2009spontaneous}.

The inability to distinguish the type of active stress from cell shape patterns in channel flow can be explained by noting the connection between the sign of the active stress and the resulting flow and cell deformation fields~\cite{nejad2025cellular}. Since the energy input and dissipation are via the active and viscous stresses respectively, balancing them gives $2\eta \mathbf{E} \sim \zeta\mathbf{Q}$ where $\eta$ is the shear viscosity and $\mathbf{E}$ is the rate of strain tensor. For extensile  systems $\zeta > 0$ and therefore $\mathbf{E} \parallel \mathbf{Q}$, i.e.~the extensional axis of the flow is along the active stress director field as shown in Figs.~\ref{fig:comparison}(e) and \siexampleturbulence(a),(c)~\cite{SI}. This straining flow deforms a cell such that its long axis extends parallel to the active stress director. On the other hand, as $\zeta < 0$ for contractile systems, $\mathbf{E} \perp \mathbf{Q}$, and the cell elongates perpendicular to the active stress director [Figs.~\ref{fig:comparison}(f) and \siexampleturbulence(b),(d)~\cite{SI}]. 

This argument underpins the cellular shape and stress profiles discussed in Fig.~\ref{fig:profiles}. In an extensile system, a splay deformation in the active stress director field induces fluid flow toward the converging end of the splay. Owing to the alignment of the shape field with the active stress director field, analogous deformations arise in the shape field, so the flow direction coincides with the converging direction of the splay deformation in the shape field. In contrast, in a contractile system, the fluid flow is directed toward the diverging end of the splay deformation in the active stress director field. However, since the shape field in this case aligns approximately perpendicular to the active stress director field, the converging direction of the resulting splay deformation in the shape field again coincides with the flow direction - mirroring the behavior observed in an extensile system. This is illustrated in Fig.~\ref{fig:comparison}(a),(b) where we plot, respectively, the angle between the active stress director and the flow $\theta_{\mathbf{Q}}$, and the angle between the shape director and the flow, $\theta_{\mathbf{R}}$. Note in particular that $\theta_{\mathbf{R}}$ is similar between the extensile and contractile systems. These results are qualitatively independent of the choice of flow aligning parameter associated with the nematic field (see Appendix A and Figs.~S2-S3 \cite{SI}).

\begin{figure}
    \includegraphics[width=8.6cm]{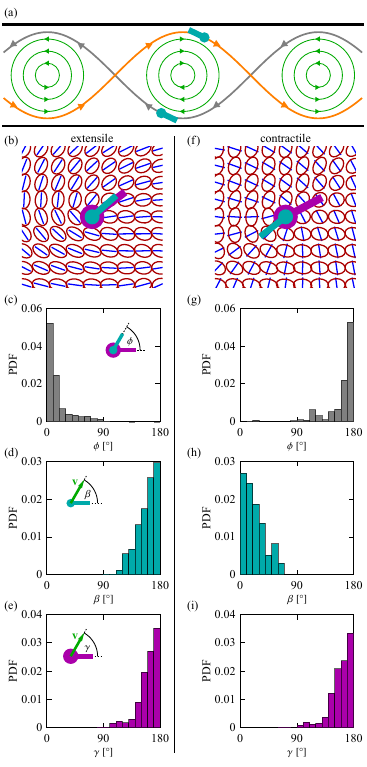}
    \caption{(a) Schematic of the dancing state, illustrating vortices in the velocity profile (green) and the trajectories (orange and gray) of $+1/2$ defects (cyan). For an extensile system: (b) Zoom on a stress defect (cyan) and its corresponding shape defect (magenta); blue lines and red ellipses show the active stress director and shape field respectively. (c) Distribution of angles between the orientation of stress defects and their nearest shape defect, $\phi$. (d) Distribution of angles between stress defects and the local velocity field, $\beta$. (e) Distribution of angles between shape defects and the local velocity field, $\gamma$. (f)-(i) Corresponding plots for a contractile system. Insets on panels (c)-(e) illustrate the relevant angles. 
    }
    \label{fig:dancing}
\end{figure}

{\textit{Dancing state}}---We next increase the activity number (defined as $N_y \times \sqrt{|\zeta|/K_\mathbf{Q}}$ where $N_y$ is the number of lattice sites along $y$ [see Appendix B]) of the channel-confined cellular layer and observe the system in its \textit{dancing} state  \cite{shendruk2017dancing}. The dancing state corresponds to a vortex-lattice in the velocity field with topological defects moving in an ordered  fashion along the channel length [Fig.~\ref{fig:dancing}(a)]. It provides a system where defect motion is regular and more easily analyzed than in active turbulence.

Topological defects are formed in both the active stress director field ($\mathbf{Q}$; stress defects) and in the shape field ($\mathbf{R}$; shape defects). As reported for active turbulence in \cite{nejad2025cellular, bera2025energy} we find that the shape defects and the stress defects are correlated in their position, orientation and trajectories. The case of an extensile and a contractile system are compared in Fig.~\ref{fig:dancing}. 

For an extensile system, both the position and the orientation of dancing $+1/2$ shape defects (magenta) coincide with those of the stress defects (cyan) [Fig.~\ref{fig:dancing}(b),(c)]. This is not surprising, as the long axis of the elongated cells align with the active stress director field. Thus, shape defects map, and are carried along with, the self-propelling $+1/2$ stress defects. Consistent with the predictions of single tensor active nematic theory, the direction of motion of the stress defects, and thus the shape defects, is tail to head~\cite{giomi2014defect}  [Fig.~\ref{fig:dancing}(d),(e)]. 

However, in the case of contractile systems, the $+1/2$ shape and stress defects coincide in their position but adopt an antiparallel orientation since the shape field is primarily orthogonal to the director field of the active stress [Fig.~\ref{fig:dancing}(f),(g)]. Therefore, while the $+1/2$ stress defects self-propel in the direction of their tails, as is expected for a contractile system in single tensor active nematic theory [Fig.~\ref{fig:dancing}(h)], the co-localized $+1/2$ shape defects move towards their heads [Fig.~\ref{fig:dancing}(i)] similar to the observations in active turbulence \cite{nejad2025cellular}.

As shape defects move from tail to head in both extensile and contractile systems, analyzing their motion does not help distinguish between the two -- mirroring the ambiguity observed in unidirectional channel flows.

\textit{Alignment at walls}---Lastly, we further increase the activity number (see Appendix B) so that the bulk of the fluid in the channel displays active turbulence, a state with highly vortical chaotic flows, in which topological defects are continually created and destroyed in pairs.  We focus on analyzing the behavior of the cells near the walls for the two types of activity, and in the absence of any thermodynamic anchoring.

\begin{figure}
    \includegraphics[width=8.2cm]{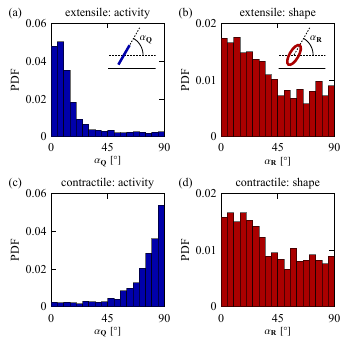}
    \caption{Distribution of angles relative to the wall tangent (shown in the insets) for (a) the active stress director field, $\alpha_{\mathbf{Q}}$ and (b) the long axis of the cell shape, $\alpha_{\mathbf{R}}$ in a turbulent field produced by extensile active stresses in a very wide channel. (c,d) Corresponding distributions in a contractile system. A snapshot of typical active stress and shape fields near the wall is shown in Fig.~\siexampleanchoring~\cite{SI}.
    }
    \label{fig:anchoring}
\end{figure}

For an extensile system, the director field of active stress aligns parallel to the wall [Fig.~\ref{fig:anchoring}(a)], whereas in a contractile system the active stress director is primarily oriented perpendicular to the wall [Fig.~\ref{fig:anchoring}(c)]. This is similar to the hydrodynamic behavior of microswimmers near rigid boundaries \cite{lauga2009hydrodynamics} or the active anchoring observed for extensile and contractile active nematics near or at an interface ~\cite{blow2014biphasic, ruske2021morphology}. Since the cell shape is extended in the direction of the active stress director field in an extensile system, the cells are elongated parallel to the walls [Fig.~\ref{fig:anchoring}(b)]. In contractile systems, because the long axis of the shape tends to orient perpendicular to the active stress director, the cell shape director also aligns parallel to the walls [Fig.~\ref{fig:anchoring}(d)].

In experiments cells are found to align parallel to the walls~\cite{duclos2018spontaneous,dong2024collective}. Our results indicate that, although the active stress director field generated by extensile and contractile activity exhibits distinct differences in orientation near the walls, the cell alignment itself cannot be used to distinguish the nature of the activity as it is similar in both cases.

\textit{Discussion}---The dynamics of epithelial monolayers has presented a persistent puzzle in active matter physics: while individual cells are intrinsically contractile due to actomyosin activity, collective tissue dynamics often resembles that of extensile active nematics. We have addressed this issue by studying a two-tensor continuum model in which tissue stresses and cell shapes are treated as distinct fields. Examining three experimentally-relevant cases, channel flows, the dynamics of topological defects, and cell orientation at rigid walls, we found that it is not possible to distinguish extensile or contractile behavior based on a simple visual inspection of cell shapes and the velocity field. Instead measurements of the stress field such as those achievable using traction force microscopy, are also needed.

Continuum theories rely on symmetry and conservation laws and hence our qualitative conclusions are insensitive to biological detail, and contribute towards interpreting the results of experiments aimed at understanding the physical nature of the forces acting within epithelial cell layers.

\begin{acknowledgments}
{\textit{Acknowledgments}}---JR and JMY acknowledge support from the UK Engineering and Physical Sciences Research Council (Award EP/W023849/1) and ERC Advanced Grant ActBio (funded as UKRI Frontier Research Grant EP/Y033981/1). JR acknowledges financial support from the Slovenian Research and Innovation Agency (development funding pillar RSF-0106 and research core funding P1-0055). SPT thanks the Royal Society and the Wolfson Foundation for the Royal Society Wolfson Fellowship award and acknowledges the support of the Department of Science and Technology, India via the research grant CRG/2023/000169. 
\end{acknowledgments}

\providecommand{\noopsort}[1]{}\providecommand{\singleletter}[1]{#1}%
%

\section{END MATTER}
\setcounter{equation}{0}
\renewcommand{\theequation}{A\arabic{equation}}
\setcounter{subsection}{0}
\setcounter{secnumdepth}{2}
\renewcommand{\thesubsection}{S\arabic{subsection}}


\textit{Appendix A: Continuum theory}---The dynamics of the cellular layer is modeled using a continuum theory with three field variables: a velocity field, $\mathbf{u}$ to describe the flow dynamics and two microstructural variables, a shape tensor $\mathbf{R}$ and a nematic order parameter $\mathbf{Q}$.

The flow dynamics is governed by the incompressible Navier-Stokes equations:
\begin{align}
\nabla\cdot\mathbf{u} &=0, \label{eqn:continuity}\\
\rho \left(\partial_t + \mathbf{u}\cdot\nabla\right) \mathbf{u} &= \nabla\cdot\boldsymbol{\Pi},\label{eqn:ns}
\end{align}
where $\rho$ is the density of the material and $\boldsymbol{\Pi}$ is the stress developed in the system. 

The microstructure of the cellular layer is encoded using two second rank tensors: a shape tensor $\mathbf{R}$ and a nematic order parameter $\mathbf{Q}$~\cite{nejad2025cellular}. The shape tensor is defined as $\mathbf{R} = \frac{1}{V}\int \mathcal{R} dV$ where $V$ is the averaging volume for the continuum approximation \cite{leal2007advanced} and $\mathcal{R} =  \sum_{i=1}^3(a_i \hat{\mathbf{r}}_i)(a_i \hat{\mathbf{r}}_i)^T$ the shape tensor for a single ellipsoidal  particle with $\hat{\mathbf{r}}_i$ a unit vector along the $i^{\textnormal{th}}$ axis of symmetry and  $a_i$ the corresponding linear dimension. Hence, the shape tensor is a symmetric quantity, its eigenvectors provide the principal directions and the square roots of its eigenvalues provide the corresponding lengths of the cells. The shape tensor evolves  according to
\begin{align}
\partial_t \mathbf{R}=& - \mathbf{u}\cdot \boldsymbol{\nabla} \mathbf{R} + \mathbf{S}_{\mathbf{R}} + \Gamma_{\mathbf{R}} \mathbf{H}_{\mathbf{R}},\label{Rdynamics}
\end{align} 
where $\partial_t \mathbf{R} + \mathbf{u}\cdot \boldsymbol{\nabla} \mathbf{R} - \mathbf{S}_{\mathbf{R}}$ is the convected derivative, 
\begin{align}
\mathbf{S}_{\mathbf{R}} =  (\xi_{\mathbf{R}} \mathbf{E} + \boldsymbol{\Omega}) \cdot \mathbf{R} + \mathbf{R}\cdot(\xi_{\mathbf{R}} \mathbf{E} - \boldsymbol{\Omega}),  
\end{align}
and $\xi_{\mathbf{R}}$ is the shape flow-aligning parameter. The rate of strain tensor $\mathbf{E}$ and the vorticity tensor $\boldsymbol{\Omega}$ are respectively  the symmetric and antisymmetric part of the velocity gradient tensor, $\nabla\mathbf{u} = \mathbf{E} - \boldsymbol{\Omega}$. Further, $\Gamma_{\mathbf{R}}$ controls the relaxation rate of the shape tensor and $\mathbf{H}_{\mathbf{R}}$, the molecular potential, in two dimensions,  is given by
\begin{align}
    \mathbf{H}_{\mathbf{R}} = \frac{k_{\mathbf{R}}}{2}\left(\frac{2\lambda_0^2}{\textnormal{Tr}(\mathbf{R})}\mathbf{I} - \mathbf{R}\right).
    \label{eqn:Rmolpot}
\end{align}
Equation~\eqref{eqn:Rmolpot} expresses the cell shape as a linear spring and is adopted from work modeling polymers in viscoelastic fluid flows \cite{beris1994thermodynamics}. $k_{\mathbf{R}}$ is the spring constant, $\lambda_0^2$ is the determinant of $\mathbf{R}$, taken as a constant in the formulation, $\textnormal{Tr}(\cdot)$ is the trace of the argument, and $\mathbf{I}$ is the identity tensor. At equilibrium, the shape tensor $\mathbf{R} = \lambda_0 \mathbf{I}$ and $\mathbf{H}_{\mathbf{R}}$ vanishes.

The second microstructural variable, the nematic order parameter $\mathbf{Q}$, is a traceless, symmetric second rank tensor that represents the magnitude and orientational order of active stresses generated in the cellular layer due to the action of molecular motors, active cytoskeletal filaments, and intercellular forces. Following the theory of active nematics \cite{simha2002hydrodynamic, doostmohammadi2018active} the governing equation for the $\mathbf{Q}$ tensor is 
\begin{align}
\partial_t \mathbf{Q}=& - \mathbf{u}\cdot \boldsymbol{\nabla} \mathbf{Q} + \mathbf{S}_{\mathbf{Q}} + \Gamma_{\mathbf{Q}} \mathbf{H}_{\mathbf{Q}},\label{qdynamics}
\end{align} 
where, as before, the convected derivative of $\mathbf{Q}$ contains
\begin{align}
    \mathbf{S}_{\mathbf{Q}} &= \left(\xi_{\mathbf{Q}}\mathbf{E}+\boldsymbol{\Omega}\right)\cdot\left(\mathbf{Q}+\frac{\mathbf{I}}{3}\right) + \left(\mathbf{Q}+\frac{\mathbf{I}}{3}\right) \cdot \left(\xi_{\mathbf{Q}}\mathbf{E}-\boldsymbol{\Omega}\right)\nonumber\\
    &-2\xi_{\mathbf{Q}} \left(\mathbf{Q}+\frac{\mathbf{I}}{3}\right) \left(\mathbf{Q}:\nabla\mathbf{u}\right),
\end{align}
where $\xi_{\mathbf{Q}}$ is the flow-aligning parameter, $\Gamma_{\mathbf{Q}} $ determines the relaxation time, and $\mathbf{H}_{\mathbf{Q}}$ is the molecular potential. $\mathbf{H}_{\mathbf{Q}}$ is also a symmetric and traceless tensor which is calculated as the variational derivative of the free energy, $\mathcal{F}_{\mathbf{Q}}$, as 
\begin{align}
    \mathbf{H}_{\mathbf{Q}} = -\frac{\delta\mathcal{F}_{\mathbf{Q}}}{\delta\mathbf{Q}} + \frac{\mathbf{I}}{3} \textnormal{Tr}\left(\frac{\delta\mathcal{F}_{\mathbf{Q}}}{\delta\mathbf{Q}}\right).
\end{align}
The free energy has a bulk and a gradient contribution, the latter arising from the nematic elasticity:
\begin{align}
    \mathcal{F}_{\mathbf{Q}} = \frac{A_0}{2}\left(1-\frac{\gamma}{3}\right)Q_{ij}^2 &- \frac{1}{3} A_0 \gamma Q_{ij}Q_{jk}Q_{ki} + \frac{1}{4} A_0\gamma (Q_{ij}^2)^2 \nonumber\\
    &+ \frac{K_{\mathbf{Q}}}{2} \left(\partial_k Q_{ij}\right)^2,
    \label{eq:FQ}
\end{align}
where $A_0$ is a phenomenological constant and $\gamma$  specifies the temperature. We use $\gamma > 3$ and hence set the temperature below the isotropic-nematic phase transition temperature \cite{chandragiri2019active}. Further, the gradient term in Eq.~\eqref{eq:FQ} assumes a single elastic constant approximation with $K_{\mathbf{Q}}$ the corresponding elastic constant.

The dynamics is driven by the total stress in Eq.~\eqref{eqn:ns}, which includes the following contributions:
\begin{enumerate}
    \item the viscous stress, 
    \begin{align}
        \boldsymbol{\Pi}^{viscous} = 2\eta\mathbf{E},
    \end{align}
    that arises from the viscosity of the material $\eta$ following a Newtonian constitutive relation.  
    \item the shape elastic stress, corresponding to Eq.~\ref{eqn:Rmolpot}
    \begin{align}
    \boldsymbol{\Pi}^{elastic,\mathbf{R}} = 2\xi_{\mathbf{R}} k_{\mathbf{R}}\left( \frac{\mathbf{R}}{\textnormal{Tr}(\mathbf{R})} - \frac{1}{2}\mathbf{I}\right),
    \label{eq:shapeelasticity}
    \end{align}
    that arises from the cell shape elasticity. The form of Eq.~\eqref{eq:shapeelasticity} originates from a linear spring approximation of the cell shape \cite{beris1994thermodynamics}. At equilibrium, the shape tensor $\mathbf{R} = \lambda_0 \mathbf{I}$ and $\boldsymbol{\Pi}^{elastic,\mathbf{R}}$ vanishes.
    \item the orientational elastic stress including the pressure field $P$,
    \begin{align}
       \boldsymbol{\Pi}^{elastic,\mathbf{Q}} &= -P \mathbf{I} + 2\xi_{\mathbf{Q}} \left(\mathbf{Q} + \frac{\mathbf{I}}{3} \right)(\mathbf{Q}:\mathbf{H}_{\mathbf{Q}}) \nonumber\\ &- \xi_{\mathbf{Q}}\mathbf{H}_{\mathbf{Q}}\cdot\left(\mathbf{Q} + \frac{\mathbf{I}}{3} \right) - \xi_{\mathbf{Q}}\left(\mathbf{Q} + \frac{\mathbf{I}}{3} \right)\cdot \mathbf{H}_{\mathbf{Q}} \nonumber\\ &- \nabla\mathbf{Q}:\frac{\delta\mathcal{F}_{\mathbf{Q}}}{\delta\nabla\mathbf{Q}} + \mathbf{Q}\cdot\mathbf{H}_{\mathbf{Q}} - \mathbf{H}_{\mathbf{Q}} \cdot \mathbf{Q},
    \end{align}
     that arises from the prevailing nematic order $\mathbf{Q}$. Any variations in $\mathbf{Q}$ from its equilibrium value generate  elastic stresses $\boldsymbol{\Pi}^{elastic,\mathbf{Q}}$ in the system and contribute to the generation of `backflow'.
    \item the active stress,
    \begin{align}
       \boldsymbol{\Pi}^{active} = -\zeta \mathbf{Q},  
    \end{align}
    that arises from the activity in the cellular layer. The proportionality constant $\zeta$ accounts for the strength of the activity, with $\zeta > 0$ and $\zeta < 0$ corresponding to extensile and contractile activity, respectively \cite{doostmohammadi2018active, doostmohammadi2021physics}.
\end{enumerate}

Hence, activity with its continuous energy input through active stress drives the system out of equilibrium. The corresponding fluid flow and flow gradients alter the shape and the nematic order parameters which further contributes to the dynamics of the cellular layer.\\

\setcounter{equation}{0}
\renewcommand{\theequation}{A\arabic{equation}}
~\\
\textit{Appendix B: Simulation details}---The theoretical model described in Appendix A was solved numerically using a hybrid lattice Boltzmann method. We used a $D3Q19$ model to implement the lattice Boltzmann algorithm. The convective-diffusive equations for the shape tensor $\mathbf{R}$ and the nematic order parameter $\mathbf{Q}$, Eqs.~\eqref{Rdynamics} - \eqref{qdynamics}, were solved using method of lines, in which the spatial derivatives are discretized using second order accurate central difference, and time integration is performed using Euler method. The simulations were performed in two dimensions by imposing periodic boundary conditions in the third dimension. Further, on open sides of the domain, periodic boundary conditions were imposed. In the case of solid walls, (i) no-slip boundary conditions for the fluid flow, (ii) Neumann boundary conditions for the shape tensor, and (iii) free anchoring boundary conditions for the nematic order parameter were used. Numerical simulations were performed till the system reached statistically steady state when the measurements were taken.

The parameters used in the simulations are as follows. The fluid has a viscosity $\eta = 0.833$ and the activity coefficient is $\zeta=\pm 0.002$. For the unidirectional flows in Figs.~\ref{fig:profiles} and \ref{fig:comparison}, we set $\xi_\textbf{Q}=0.9$,  $k_\textbf{R}=10^{-4}$, and $K_\textbf{Q}=4\times10^{-3}$ ($\xi_\textbf{Q}=0.1$,  $k_\textbf{R}=10^{-4}$, and $K_\textbf{Q}=10^{-2}$) for the extensile (contractile) system. For the dancing state in Fig.~\ref{fig:dancing}, we set $\xi_\textbf{Q}=0.9$,  $k_\textbf{R}=6\times10^{-4}$, and $K_\textbf{Q}=2\times10^{-3}$ ($\xi_\textbf{Q}=-0.9$,  $k_\textbf{R}=5\times10^{-4}$, and $K_\textbf{Q}=2\times10^{-3}$) for the extensile (contractile) system. To study active turbulent flows at walls in Fig.~\ref{fig:anchoring}, we set $\xi_\textbf{Q}=0.1$,  $k_\textbf{R}=5.5\times10^{-4}$, and $K_\textbf{Q}=10^{-3}$ for both systems. In the free energy for the $\mathbf{Q}$ tensor, $A_0 = 0.0843$ and $\gamma = 3.085$, and in that for $\mathbf{R}$, $\lambda_0 = 0.1283$. The flow aligning parameter for the shape tensor is set to $\xi_{\mathbf{R}} = 1$.  The unidirectional flow and dancing state simulations have channel length $N_x = 96$ and width $N_y = 16$; the active turbulent flows have channel length and width $N_x = N_y = 512$. Therefore, the activity number, defined as $N_y \times \sqrt{|\zeta|/K_\mathbf{Q}}$, varies from 7.16 to 724 in our simulations. 

\end{document}


\preprint{}

\title{
\textbf{Supplemental Material: \\ Why Extensile and Contractile Tissues Could be Hard to Tell Apart}}
\author{Jan Rozman}
\affiliation{Rudolf Peierls Centre for Theoretical Physics, University of Oxford, Oxford OX1 3PU, United Kingdom}
\affiliation{Faculty of Mathematics and Physics, University of Ljubljana, Jadranska 19, SI-1000 Ljubljana, Slovenia}
\affiliation{ Jo\v zef Stefan Institute, Jamova 39, SI-1000 Ljubljana, Slovenia}

\author{Sumesh P. Thampi}%
\affiliation{Rudolf Peierls Centre for Theoretical Physics, University of Oxford, Oxford OX1 3PU, United Kingdom}
\affiliation{Department of Chemical Engineering, Indian Institute of Technology, Madras, Chennai, India 600036}

\author{Julia M. Yeomans}%
\affiliation{Rudolf Peierls Centre for Theoretical Physics, University of Oxford, Oxford OX1 3PU, United Kingdom}

\date{\today}

{
\let\clearpage\relax
\maketitle
}

\section*{Defect detection}

To identify defects in the shape field, we first define for each lattice site a shape director 
\begin{equation}
    \textbf{n}_\textbf{R}=\left[\cos\left(\nu_\textbf{R}\right),\sin\left(\nu_\textbf{R}\right)\right],
\end{equation}
where $\nu_\textbf{R}$ is an angle giving the direction of the shape long axis at that site relative to the $x$ axis. From this, we construct a modified shape field
\begin{equation}
    \tilde{\textbf{R}}=\textbf{n}_\textbf{R}\otimes \textbf{n}_\textbf{R}-\frac{1}{2}\textbf{I},
\end{equation}
where $\textbf{I}$ is the identity tensor. The angle giving the direction of the active stress field director relative to the $x$ axis is denoted $\nu_\textbf{Q}$. Both $\nu_\textbf{Q}$ and $\nu_\textbf{R}$ are confined to the range $\left[-\pi/2,\pi/2\right]$.

We determine the position of defects by calculating the winding number~\cite{huterer2005distribution,saw2017topological}, following the approach described in Ref.~\cite{killeen2022polar}. The winding number of a lattice site is defined as
\begin{equation}
    \Delta \nu_\textbf{W} = \sum_l \delta \nu_{\textbf{W},(l,l+1)},
\end{equation}
where the sum is over the eight neighbors $l$ of that lattice site in counterclockwise order and $\delta \nu_{\textbf{W},(l,l+1)}$ (for $\mathbf{W} \in \{\mathbf{Q}, \mathbf{R}\}$) is the change in the director/long axis angle between sites $l$ and $l+1$, given as
\begin{equation}
    \delta \nu_{\textbf{W},(l,l+1)} = \nu_{\textbf{W},l+1}-\nu_{\textbf{W},l}+\rho
\end{equation}
where $\nu_{\textbf{W},l}$ is the angle at site $l$ and 
\begin{align}
    \rho=0\ &&\textrm{if}\ &&| \nu_{\textbf{W},l+1}- \nu_{\textbf{W},l}|\leq \pi/2,\\
    \rho=\pi\ &&\textrm{if}\ && \nu_{\textbf{W},l+1}- \nu_{\textbf{W},l} < -\pi/2,\\
    \rho=-\pi\ &&\textrm{if}\ && \nu_{\textbf{W},l+1}- \nu_{\textbf{W},l} > \pi/2.
\end{align}
The defect charge is then equal to
\begin{equation}
    k=\frac{\Delta \nu_\textbf{W}}{2\pi}.
\end{equation}
Defects are only detected on every other point to prevent double counting from overlapping winding number loops. 
    
The direction of the defect is calculated following Ref.~\cite{vromans2016orientational},
\begin{equation}\label{eq:direction}
    \psi=\frac{k}{1-k}\arctan\left[\frac{\textrm{sgn}\left(k\right)\partial_x W_{xy}-\partial_y W_{xx}}{\partial_x W_{xx} + \textrm{sgn}\left(k\right)\partial_y W_{xy}} \right],
\end{equation}
where $\mathbf{W} \in \{\mathbf{Q}, \tilde{\mathbf{R}}\}$. $\psi$ is the angle relative to the $x$ axis giving the direction from a $+1/2$ defect's core to its tail or from the core of a $-1/2$ defect to one of its three ``arms''. The derivatives in Eq.~\eqref{eq:direction} are taken as the difference between values of the tensor component at the two nearest neighbors along the direction in which the derivative is taken. Note that we do not calculate defect charges for the rows immediately adjacent to the walls.

\newpage
\bibliographystyle{apsrev4-1}
\providecommand{\noopsort}[1]{}\providecommand{\singleletter}[1]{#1}%

\newpage

\section*{Supplementary figures}
\begin{figure}[h]
    \centering
    \includegraphics[width=16 cm]{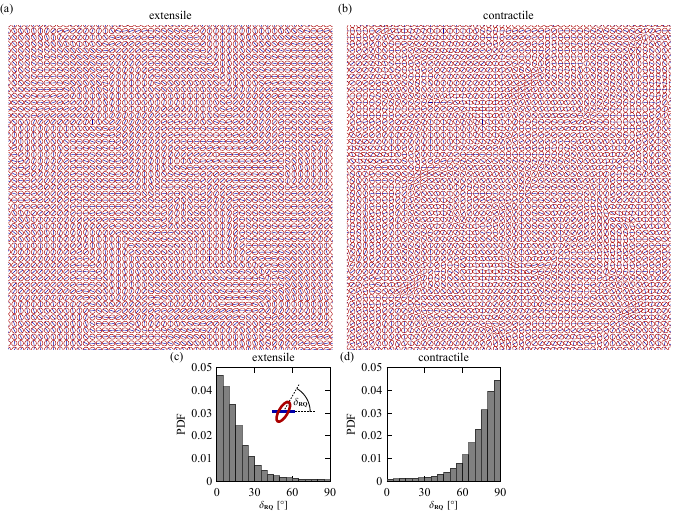}
    \caption{(a,b) Example of a part of a system with periodic boundary conditions for extensile (a) and contractile (b) activity. Blue lines show the active stress directors and red ellipses show the shape field. (c,d) Corresponding distribution of angles between the active stress director and the shape long axis in the extensile (c) and contractile (d) system. The parameters are $N_x=N_y=256$, $\xi_\textbf{Q}=0.1$, $K_\textbf{Q}=10^{-3}$, and $k_\textbf{R}=10^{-3}$ ($k_\textbf{R}=5\times10^{-4}$) for the extensile (contractile) system.}
    \label{sfig:turbulence}
\end{figure}

\begin{figure}[h]
    \centering
    \includegraphics[width=12 cm]{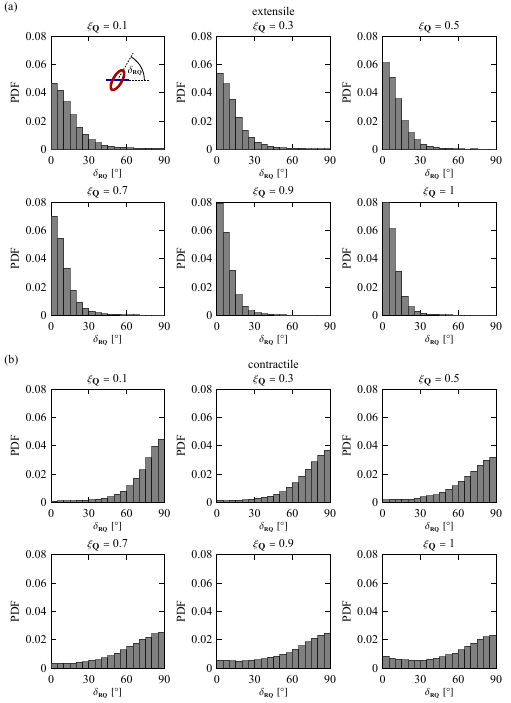}
    \caption{Probability distribution of $\delta_{\mathbf{R}\mathbf{Q}}$, the angle between active stress director and the cell shape long axis, for an unconfined, periodic system at different values of the flow aligning parameter $\xi_{\mathbf{Q}}$ for~(a) an extensile and~(b) a contractile system. The inset on panel (a) illustrates the definition of the angle $\delta_{\mathbf{R}\mathbf{Q}}$. Parameters as in Fig.~\ref{sfig:turbulence}.
    }
\end{figure}

\begin{figure}[h]
    \centering
    \includegraphics[width=16 cm]{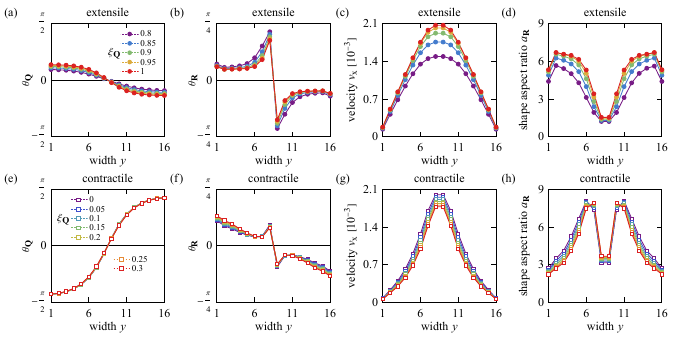}
    \caption{Comparison of channel flows for different values of the flow aligning parameter $\xi_{\mathbf{Q}}$ in an extensile (a-d) and a contractile (e-h) system: profiles of (a,e) $\theta_{\mathbf{Q}}$, the angle of the active stress director relative to the flow direction, (b,f) $\theta_{\mathbf{R}}$, the shape long axis angle relative to the flow direction, (c,g) velocity, and (d,h) shape aspect ratio.
    }
\end{figure}

\begin{figure}[h]
    \centering
    \includegraphics[width=16cm]{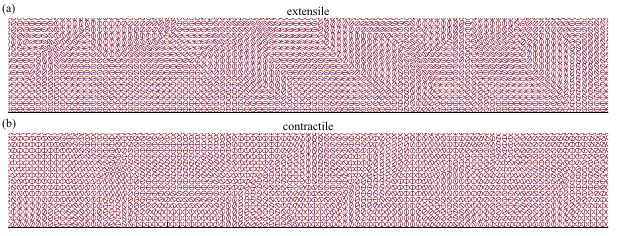}
    \caption{Example of cell orientations at the wall (black line) for a system in a 512-wide channel  in which the bulk of the fluid exhibits active turbulence for extensile (a) and contractile (b) activity. Blue lines show the active stress directors and red ellipses show the shape field.
    }
    \label{sfig:anchoring}
\end{figure}